\documentstyle[12pt]{article}
\def\be{\begin{equation}}\def\ee{\end{equation}}

\newcommand{\p}[1]{(\ref{#1})}
\begin{document}
\renewcommand{\thefootnote}{\arabic{footnote}}
\begin{flushright}
HUB-EP-98/66\\
DFPD 98/TH/45\\
hep-th/9809213\\
Phys. Lett. B (in press)
\end{flushright}

\vspace{1truecm}
\begin{center}
{\large\bf 
On gauge-fixed superbrane actions in AdS superbackgrounds}

\bigskip
Paolo Pasti${}^*$, Dmitri~Sorokin ${}^{**}$\footnote{
Alexander von Humboldt fellow.\\
On leave from Kharkov Institute of
Physics and Technology, Kharkov, 310108, Ukraine.} 
and Mario Tonin$^*$

\bigskip
${}^*$ Universit\`a Degli Studi Di Padova,
Dipartimento Di Fisica ``Galileo Galilei''\\
ed INFN, Sezione Di Padova\\
Via F. Marzolo, 8, 35131 Padova, Italia\\

\bigskip
${}^{**}$ Humboldt-Universit\"at zu Berlin\\
Mathematisch-Naturwissenshaftliche Fakultat\\
Institut f\"ur Physik\\
Invalidenstrasse 110, D-10115 Berlin, Germany

\bigskip
{\bf Abstract}
\end{center}
To construct actions for describing superbranes propagating in $AdS\times S$
superbackgrounds
we propose a coset space realization of these superbackgrounds
which results in a short polynomial fermionic dependence (up to the sixth power in
Grassmann coordinates) of target superspace supervielbeins and superconnections.
 Gauge fixing $\kappa$--symmetry in a way compatible
with a static brane solution further reduces the fermionic dependence down 
to the fourth power. Subtleties of consistent gauge fixing 
worldvolume diffeomorphisms and $\kappa$--symmetry of the superbrane actions
are discussed.



\newpage
Worldvolume dynamics of super--p--branes (M-branes and D-branes)
propagating in 
Anti--de--Sitter supergravity backgrounds has been understood to 
give rise to  $(p+1)$--dimensional superconformal theories 
\cite{Maldacena}--\cite{Witten}. In connection with this observation
an extensive 
study of superbrane worldvolume actions in these superbackgrounds 
has been carried out \cite{CKVP}--\cite{P1}.

To reduce the worldvolume action of the superbrane to that of
the superconformal theory one should 
know an explicit form of the geometrical quantities which describe
the AdS background (supervielbeins, superconnections and antisymmetric
gauge superfields), to gauge fix worldvolume
diffeomorphisms for bosonic target--space coordinates of the superbrane,
and to eliminate half the fermionic target--space coordinates by gauge 
fixing kappa--symmetry of the superbrane worldvolume action.

These problems are mostly technical ones, and to arrive at their simplest
solution one, for instance, can try to 
guess an appropriate realization of the AdS
target superspace as a coset superspace $K=G/H$, where $G$ is the isometry
supergroup of the background and $H$ is it's isotropy (or stability) 
subgroup. For these purposes the coset--space techniques has been developed
in a number of papers \cite{MT}--\cite{DFFFTT,dWPPS}. 
Another possibility is to directly solve supergravity constraints \cite{Claus}.  
Such realizations 
produce a complicated form of the target superforms with the dependence on
Grassmann coordinates $\Theta$ up to a 32--nd power.
Upon imposing a so called Killing spinor \cite{RKILL} or supersolvable algebra
\cite{DFFFTT} condition for gauge fixing the $\kappa$--symmetry, a significant
simplification of the structure of the superforms was achieved, so that
only the second power in $\Theta$ remains in the supervielbeins and
superconnections. The choice of a gauge fixing condition is a subtle point
in the sense that it must be compatible with the classical solution of
brane equations of motion chosen as a vacuum over which a perturbative
theory is developed. As we shall see the gauge fixing condition mentioned
above is not always compatible with a static brane solution embedded into
the AdS part of the target superspace. In these cases another
$\kappa$--symmetry gauge fixing is required.

In this paper we propose a coset space realization of AdS superbackgrounds
which results in a short polynomial fermionic dependence (up to the sixth power in
$\Theta$) of the superforms even before fixing the $\kappa$--symmetry.
Gauge fixing the $\kappa$--symmetry in a way consistent with the static
brane solution further reduces $\Theta$--dependence down to the fourth
power.

The results
obtained are generic for the cases of a D3-brane in IIB D=10
supergravity, and an M2--brane and an M5--brane in D=11 supergravity,
which are considered
simultaneously. They should also be applicable to a superstring propagating
in $AdS^3\times S^3$.

The general form of the superbrane action is (details can be found in 
\cite{BST,dbrane,5b})
\begin{equation}\label{action}
S=I_0+I_{WZ},
\end{equation}
where
\begin{equation}\label{I0}
I_0=-\int_{{\cal M}_{p+1}}d\xi^{p+1}\sqrt{-detg_{ij}}+ ...
\end{equation}
is a Nambu--Goto or Dirac--Born--Infeld--type part of the action (the dots
stand for the contribution of worldvolume gauge fields, if present),
\begin{equation}\label{g}
g_{ij}(\xi)=\partial_iZ^ME_M^{\hat a}E_{\hat aN}\partial_jZ^N
\quad (i,j=0,1...,p)
\end{equation}
is an induced worldvolume metric and $E_M^{\hat a}(Z), E_M^{\hat \alpha}(Z)$
are  supervielbeins of target superspace parametrized by the coordinates
$Z^M=(X^{\hat m}, \Theta^{\hat\alpha})$.
\begin{equation}\label{IWZ}
I_{WZ}=-\int_{{\cal M}_{p+1}} A^{(p+1)}+...
\end{equation}
is the Wess-Zumino part of the superbrane action whose leading term is the
worldvolume integral over the pullback of a target--space 
$(p+1)$--form gauge superfield.
It is convenient to take the following metric as 
the bosonic part of the $AdS^{p+2}\times S^{D-p-2}$ superbackground metric
\begin{equation}\label{metric}
ds^2 = \left({r\over R}\right)^{{2(D-p-3)}\over{p+1}} dx^i \eta_{ij} dx^j 
+ \left(\frac Rr\right)^2 dr^2 + R^2 d\Omega^2\,,
\end{equation}
where $x^i,r$ $(i=0,1,...,p)$ are coordinates of the AdS and $d\Omega^2$
stands for a  metric of the sphere of a radius $R$ parametrized by coordinates
$y^{a'}$ $(a'=1',...,(D-p-2)')$.

The nonvanishing pure 
bosonic components of the gauge field $A^{(p+1)}$ in the 
$AdS\times S$ background are
\begin{equation}\label{A}
A^{(p+1)}=dx^p...dx^0({r\over R})^{D-p-3} + O(\Theta).
\end{equation}

The action \p{action} is invariant under the fermionic $\kappa$--symmetry
transformations
\begin{equation}\label{kappa}
\delta_\kappa Z^ME_M^{\hat a}=0, \quad 
\delta_\kappa Z^ME_M^{\hat \alpha}=\kappa^{\hat\beta}(\xi)
(1+\bar\Gamma)_{\hat\beta}^{~\hat\alpha},
\end{equation}
(plus a variation of worldvolume gauge fields)
where $1+\bar\Gamma$ is a spinor projector. The form of $\bar\Gamma$
is specific for each superbrane \cite{BST,dbrane,5b} 
but for every brane the leading term
of  $\bar\Gamma$ is
\begin{equation}\label{ga}
\bar\Gamma={1\over{(p+1)!\sqrt{-g}}}\varepsilon^{i_1...i_{p+1}}
\Gamma_{i_1...i_{p+1}}+ ...
\end{equation}
where $\Gamma_{i_1...i_{p+1}}$ is the antisymmetric product of target--space
gamma--matrices pulled back on to the worldvolume, i.e. 
$\Gamma_{i}=\partial_iZ^ME_M^{\hat a}\Gamma_{\hat a}$.

Note that a classical static ``vacuum'' solution of the superbrane 
equations of motion which follow
from \p{action} is
\begin{equation}\label{vac}
\xi^i=x^i,\quad r=const, \quad y^{a'}=const, \quad \Theta=0, 
\end{equation}
and the worldvolume gauge fields are zero.

For this static solution the action \p{action} vanishes, which is called
the no--force
condition, since there is no a potential which would push the brane to
a boundary of AdS \cite{end,CKKTVP,DFFFTT}. 
This, in particular, means that the static gauge
$\xi^i=x^i$ for the worldvolume diffeomorphisms is compatible with the vacuum
solution. 

In the case of the static 
vacuum solution the $\kappa$--symmetry projector takes a
simple form
\begin{equation}\label{g0}
1+\bar\Gamma=1+\bar\gamma,
\end{equation}
where
$\bar\gamma=\Gamma^{01...p}$ in the case of the M2 and the M5--branes,
and in the case of the IIB D3-brane 
$\bar\gamma=\epsilon^{IJ}\Gamma^{01...3}$ (where $I,J=1,2$ number the
two D=10 Majorana--Weyl spinors). 

Note that if we chose another sign in front of the Wess--Zumino term,
the sign in the $\kappa$--symmetry projector \p{kappa} would also change.
In this case, with the same choice of \p{metric} and \p{A}, 
the no-force condition would be satisfied if in the 
solution \p{vac} one of the worldvolume coordinates is equal to minus the 
corresponding AdS coordinate (and the static gauge must also respect this
minus sign). At this the
vacuum value of the 
$\kappa$--symmetry projector would remain the same as in \p{g0}.
This is important for consistent gauge fixing the kappa--symmetry.

Let us now turn to the description of the AdS superbackground.
The isometry supergroup of the target superspace in question
is $G=OSp(8|4)$, $SU(2,2|4)$ or $OSp(2,6|4)$ 
(which corresponds, respectively, to an M2--brane, D3--brane or an M5--brane)
whose bosonic subgroup is $SO(2,p+1)\times SO(D-p-1)$ ($p=2,3,5;$ $D=10,11$)
and the stability subgroup
is $H=SO(1,p+1)\times SO(D-p-2)$. 
As a homogeneous coset superspace the target 
superspace is realised as $K=G/H$, and
its bosonic subspace parametrized by coordinates $X^{\hat m}$ ($\hat 
m=0,1,...,D$) is 
$AdS^{p+2}\times S^{D-p-2}=
{{SO(2,p+1)\times SO(D-p-1)}\over{SO(1,p+1)\times SO(D-p-2)}}$.
Supervielbeins $E^{\hat a}(X,\Theta)$, $E^{\hat\alpha}(X,\Theta)$
and a superconnection $\Omega^{\hat a\hat b}$ defining the geometry of $K$ 
are determined as components of the Cartan one--form
\begin{equation}\label{1}
K^{-1}dK=E^{\hat a}P_{\hat a}+E^{\hat\alpha}\hat Q_{\hat\alpha}
+\Omega^{\hat a\hat b}J_{\hat a\hat b},
\end{equation}
where $P_{\hat a}$, $\hat Q_{\hat\alpha}$ and $J_{\hat a\hat b}$ are the 
generators of $G$, $J_{\hat a\hat b}$ are the generators of the stability
subgroup 
$H=SO(1,p+1)\times SO(D-p-2)$, 
and $P_{\hat a}$ and $\hat Q_{\hat\alpha}$ are, 
respectively, the bosonic and fermionic generators which correspond to 
the coset $K=G/H$. The explicit form of the algebra of these generators
can be found in \cite{MT,KRR,MT2,dWPPS}.

With respect to $SO(1,p+1)\times SO(D-p-2)$ the bosonic 
generators of $G$ split as follows.

$P_{\hat a}\equiv (P_a,P_{a'})$, $J_{\hat a\hat b}\equiv$($J_{a b}$,
$J_{a'b'})$, where $a=0,1,...,p+1$ and $a'=1',...,(D-p-2)'$ are indices
of $AdS^{p+2}$ and $S^{D-p-2}$ coordinates, respectively. 

$P_a$ and $P_{a'}$ are generators of the cosets 
${{SO(2,p+1)}\over{SO(1,p+1)}}$ and ${{SO(D-p-1)}\over{SO(D-p-2)}}$, 
respectively, and
$
\hat Q_{\hat\alpha}
$
is a 32--component spinor generator which is a Majorana spinor in
$D=11$, or consists of two Majorana--Weyl spinors in $D=10$. 

Our goal is to find the most suitable realization of the coset 
$K=G/H$ for getting (upon gauge fixing worldvolume diffeomorphisms and 
$\kappa$--symmetry) the simplest form of the worldvolume
pullbacks of the supervielbeins and the superconnection. 

A suitable realization turns out to be the one which corresponds to a manifest
superconformal structure of the supergroup $G$ in the $p+1$-dimensional
worldvolume of the corresponding superbrane, namely  
\begin{equation}\label{3}
K=e^{{x^i}\Pi_i}e^{\rho D}e^{y^{a'}P_{a'}}
e^{\eta^{\hat\alpha}Q_{\hat\alpha}}e^{\theta^{\hat\alpha}S_{\hat\alpha}},
\end{equation}
where (as above) $x^i$ $(i=0,...,p)$ and $\rho={{D-p-3}\over{p+1}}\log{r\over 
R}$
are coordinates of
$AdS^{p+2}$, 
$y^{a'}$ $(a'=1',...,(D-p-2)')$ 
are coordinates of $S^{D-p-2}$, 
\begin{equation}\label{pi}
\Pi_i =P_i+J_{ip+1}
\end{equation}
is the momentum generator,
\begin{equation}\label{D}
D=P_{p+2} 
\end{equation}
is the dilatation generator
of conformal transformations in $d=p+1$ worldvolume and
\begin{equation}\label{K}
K_i=P_i-J_{ip+1}
\end{equation} 
are special conformal transformations.

$\theta$ and $\eta$ are the following projections
of $\Theta$--coordinates\footnote{Note that 
there always exists a realization of
the gamma-matrices in which $\bar\gamma$ is symmetric.}
\begin{equation}\label{5}
\theta=\Theta(1-\bar\gamma)=(1-\bar\gamma)\Theta, \qquad
\eta=\Theta(1+\bar\gamma)=(1+\bar\gamma)\Theta.
\end{equation}
And
\begin{equation}\label{6}
Q={1\over 2}(1+\bar\gamma)\hat Q, \qquad
S={1\over 2}(1-\bar\gamma)\hat Q. 
\end{equation}
$\bar\gamma$ is the same as in \p{g0}, i.e. it coincides
with the classical vacuum value of the matrix $\bar\Gamma$ in the
$\kappa$--symmetry projector \p{kappa}, \p{ga} of the superbrane, 
which corresponds
to the static brane configuration \p{vac}. 

It is important to notice that the projector $(1-\bar\gamma)$ also appears
in the form of  the Killing spinors on $AdS\times S$ 
\cite{LPR,RKILL}
\begin{equation}\label{kil}
\epsilon_{Kill}=\phi(r,y^{a'})[1-{1\over
2}x^i\Gamma_i\Gamma_r(1-\bar\gamma)]\epsilon_{const},
\end{equation}
where $\phi(r,y^{a'})$ is a spinor matrix.
At the same time notice that it is the Grassmann coordinate $\theta$ which is
constructed
by applying to $\Theta$ the projector $1-\bar\gamma$. 
This means that in the case of the classical static 
solution \p{vac}
$\eta$ transforms under the $\kappa$--symmetry transformations, while
$\theta$ remains invariant. Therefore, $\eta$ corresponds to the Killing
spinor in the bulk which has the property that
$(1-\bar\gamma)\epsilon^{Bulk}_{Kill}=0$ \cite{RKILL}.

The generators discussed above plus special conformal boosts $K_i$, the boosts
$P_{a'}$ on the $S^{D-p-2}$ sphere and 
generators $J_{ij}$ and $J_{a'b'}$ of $SO(1,p)\times SO(D-p-2)$ form
the algebra of superconformal transformations in the
$d=p+1$ worldvolume of the superbrane.

The advantages of the realization \p{3} are the following
\footnote{An analogous nonlinear realization of the $SU(2,2|N)$ superconformal
group in $D=4$ (relevant to the D3--brane) was constructed a long time ago
in \cite{abz}.
For the M2--brane case a similar realization has been
considered in \cite{DFFFTT}, but the choice of the supersolvable algebra gauge
caused a brane configuration studied therein to live only on a
boundary of $AdS^4$.}:

i) At $\theta=\eta=0$ it
directly leads to the $AdS^{p+2}$ metric in the form \p{metric}.
This choice of the AdS metric allows one to gauge fix
worldvolume diffeomorphisms by identifying $x^i$ coordinates of the
$AdS^{p+2}$ with coordinates $\xi^i$ of the $d=p+1$ worldvolume, i.e.
to choose
the static gauge. As we have already discussed above the choice of the
static gauge must be in agreement with the no--force condition.

ii) In \p{3} used is the splitting of $\hat Q=(Q,S)$ defined in
\p{6} with $Q$ and $S$ satisfying simple commutation relations with
$\Pi_i$, $K_i$ and $D$
$$
\{Q,Q\}\sim \Gamma^i\Pi_i, \quad [\Pi_i,Q]=0, \quad
\{S,S\}\sim \Gamma^iK_i, \quad  [K_i,S]=0, 
$$
\begin{equation}\label{com}
[D,Q]=Q, \quad [D,S]=-S,\quad
[\Pi_i,S]\sim \Gamma_iQ, \quad [K_i,Q]\sim \Gamma_iS.
\end{equation}
As we will see below, this allows one to get in a quite
simple way and in a closed (explicit) form the $\theta$-- and 
$\eta$--dependence of the supervielbeins and superconnections.  

iii) Conformal supersymmetry generated by $S_{\hat\alpha}$ is nonlinearly 
realized on the worldvolume fields, this implies that conformal
supersymmetry is spontaneously broken and the coordinates
$\theta(\xi)$ are Goldstone fermionic fields  which 
are part of the physical spectrum of the $d=p+1$ superconformal theory
in the worldvolume.
The fermionic coordinates $\eta(\xi)$ correspond to unbroken 
supersymmetries generated by $Q_{\hat\alpha}$. They can be gauge fixed
to zero by means of $\kappa$--symmetry transformations. As we have discussed
the consistency of such a gauge choice is ensured by the presence of 
the ``vacuum" $\kappa$--symmetry projector in the definition of $\eta$
\p{5}.

Using the methods developed in \cite{MT} we get the following
general form of the Cartan forms $K^{-1}dK$ (with $K$ defined in \p{3}):
\begin{equation}\label{kdk}
K^{-1}dK=e^i\Pi_i+e^{\hat\alpha}Q_{\hat\alpha}
+f^iK_i+f^{\hat\alpha}S_{\alpha}+E^AT_A,
\end{equation}
where $T_A$ stand for the generators $D$, $J_{ij}$, $P_{a'}$ and 
$J_{a'b'}$, 
\begin{equation}\label{ei}
e^i=e^i_0(x,\rho)+{\Delta\eta}\Gamma^i\eta, \quad \Delta\eta \equiv
d\eta+e^A_0\eta
t_A;
\end{equation}
\begin{equation}\label{eq}
e^{\hat\alpha}=(\Delta\eta+e^i\theta\Gamma_i)^{\hat\alpha}, 
\end{equation}
\begin{equation}\label{es}
f^{\hat\alpha}=\big[\Delta\theta+(\Delta\eta h^A\theta)\theta g_A+
e^i(\theta\Gamma_ih^A\theta)\theta g_A\big]^{\hat\alpha}, 
\quad \Delta\theta\equiv d\theta+E^A\theta g_A,
\end{equation}
\begin{equation}\label{EA}
E^A=e^A_0(x,\rho,y)+\Delta\eta h^A\theta+e^i(\theta\Gamma_i h^A\theta),
\end{equation}
\begin{equation}\label{fi}
f^i=\Delta\theta\Gamma^i\theta+(\Delta\eta h^A\theta)
(\theta g_A\Gamma^i\theta)+e^j(\theta\Gamma_jh^A\theta)(\theta
g_A\Gamma^i\theta), 
\end{equation}
where $e^i_0(x^i,\rho)$ are bosonic vielbeins of $AdS^{p+2}$,
$e^A_0(x^i,\rho,y^{a'})$ are
bosonic vielbeins and connections corresponding to $T_A$, and
$t^{~~\hat \beta}_{A\hat\alpha}$, $h^A_{\hat\alpha\hat\beta}$ and $g^{~~\hat 
\beta}_{A\hat\alpha}$ are the following
structure constants of the superalgebra $G$
\begin{equation}\label{9}
[T_A,Q_{\hat\alpha}]=t^{~~\hat \beta}_{A\hat\alpha}Q_{\hat\beta}, \quad
\{Q_{\hat\alpha},S_{\hat\beta}\}=h^A_{\hat\alpha\hat\beta}T_A, \quad
[T_A,S_{\hat\alpha}]=g^{~~\hat \beta}_{A\hat\alpha}S_{\hat\beta}.
\end{equation}
Upon gauge fixing $\kappa$--symmetry by putting $\eta=0$, the 
supervielbeins and superconnections take a simpler form
\begin{equation}\label{7}
K^{-1}dK|_{\eta=0}=e^i_0\Pi_i+e^{\hat\alpha}_0Q_{\hat\alpha}
+f^i_0K_i+f^{\hat\alpha}_0S_{\alpha}+E^A_0T_A,
\end{equation}
\begin{equation}\label{8}
e^{\hat\alpha}_0=e^i_0(\theta\Gamma_i)^{\hat\alpha},
\qquad E^A_0=e^A_0+e^i_0(\theta\Gamma_ih^A\theta),
\end{equation}
$$
f^{\hat\alpha}_0=
d\theta^{\hat\alpha}+E^A_0(\theta g_A)^{\hat\alpha}+
e^i_0(\theta\Gamma_ih^A\theta)(\theta g_A)^{\hat\alpha}
=\Delta\theta^{\hat\alpha}+e^i_0(\theta\Gamma_ih^A\theta)(g_A\theta)^{\hat\alpha
},
$$
\begin{equation}\label{10}
f^i_0=\Delta\theta\Gamma^i\theta+e^j_0(\theta\Gamma_jh^A\theta)
(\theta g_A\Gamma^i\theta).
\end{equation}
In eqs. \p{ei}--\p{10} the exact coefficients in the terms of the superforms
depend on the superalgebra of $G$ (and in each case one may expect further
simplification of eqs. \p{eq}--\p{fi}).
We observe that upon the gauge fixing ($\eta=0$) 
the fourth power is the maximum power of $\theta$ in the 
definition of the supervielbeins and the superconnections.

The bosonic $AdS^{p+2}$ vielbeins $e^i_0$ which appear in the realization 
under consideration correspond to the choice \p{metric} of the $AdS\times S$ 
metric. Note also that the following combinations
of supervielbeins contribute into the definition of the induced
worldvolume metric \p{g}
\begin{equation}\label{11}
E_M^{\hat a}=(e^i_M+f^i_M, E^\rho_M, E^{a'}_M).
\end{equation}
This is because the indices ${\hat a}$ in \p{g} correspond to the AdS boost 
generators $P^i$ (eq. \p{3}), and not to $\Pi^i$ (eq. \p{pi}). 
 
 Substituting the gauge fixed values of the supervielbeins into the 
superbrane action one gets an action for an interacting
superconformal field theory in $d=p+1$, whose structure remains rather
complicated and contains interacting terms up to an 8--th power in
$\theta$.

\bigskip
{\bf Discussion}

We have considered a coset realization of the $AdS\times S$  target
superspaces which, upon gauge fixing the $\kappa$--symmetry, allows one 
 to simplify to a certain extent the form of superbrane actions in these
backgrounds.

The $\kappa$--symmetry gauge fixing considered in this paper should reduce
the $\Theta$--dependence of target superforms to the fourth power also
in the realizations of the AdS superspaces considered in 
\cite{MT,KRR,dWPPS,Claus}.

It would be of interest to establish the relationship between different
realizations of the supervielbeins and superconnections before fixing
$\kappa$--symmetry. It may happen that in the realizations of 
\cite{MT,KRR,dWPPS,Claus}, the
power of $\Theta$ is in fact less than 32 due to some matrix identities.
Or the superforms of different realizations can be related by transformations
(such as super--Weyl transformations) under which the supergravity 
constraints are invariant. 

Much simpler form of the supervielbeins and superconnections (up to the
second power in $\theta$) might be obtained if in \p{ei}--\p{fi} 
instead of the
$\eta$--coordinate it might be possible to put to zero the
$\theta$--coordinate. Such a gauge choice would correspond to the Killing
spinor gauge of \cite{RKILL} or to a supersolvable algebra of \cite{DFFFTT}.
In this gauge the static ``vacuum" value of the $\kappa$--symmetry
projector would be $(1-\bar\gamma)$ and not  $(1+\bar\gamma)$ as in
eq. \p{g0}. This would correspond to an
``anti"--static gauge with respect to the worldvolume diffeomorphisms, i.e.
when one of the worldvolume coordinates is identified with minus 
the corresponding coordinate of $AdS$ (for example, identify the time
coordinates as $\xi^0=-x^0$). 
However, there is no a classical static solution
of the
superbrane equations of motion obtained from the action \p{action} 
in the background \p{metric}, \p{A} which would
be
compatible with the ``anti"--static gauge unless the $AdS$ radial
coordinate $r$ is zero. 

Alternatively, one might change the sign of the Wess--Zumino term keeping
the form \p{metric}, \p{A} of the background. This would correspond
to an anti-brane of an opposite charge 
propagating in the AdS background generated by
a large number of branes put on the top of each other, and it is known that
there is a force between branes and anti-branes.

Thus, the no--force condition would not hold and
the Killing spinor (or supersolvable algebra) gauge is incompatible with
the static solution in
these cases. It was
admissible in the case of a IIB superstring in $AdS^5\times S^5$ 
\cite{P,KR,KT}
since the superstring $\kappa$--symmetry projector differs from
$\bar\Gamma$ in eqs. \p{kappa} and \p{ga} for the D3--brane projector, 
and the no--force problem and the consistency of the gauge choice 
should be studied there from a different point of view (as discussed, 
for example, in \cite{KT}).

A group--theoretical and geometrical 
reason why in the cases considered in this paper
the $\kappa$--symmetry allows one to eliminate $\eta(\xi)$ and not
$\theta(\xi)$ is that $\theta(\xi)$ are worldvolume
Goldstone fields of spontaneously
broken special conformal supersymmetry in the bulk, 
while $\eta$ correspond to
unbroken worldvolume supersymmetry which in the superembedding approach
to  describing superbranes 
has been known to be an irreducible realization of the $\kappa$--symmetry
\cite{stv,bpstv,hsw} (for recent reviews see \cite{rev}). 
In this approach $\eta$ can be identified with 
Grassmann coordinates which parametrize worldvolume supersurface embedded 
into target superspace. Putting the worldvolume Grassmann coordinates to zero
one reduces the superembedding models to the Green--Schwarz formulation of the
superbranes.

All this does not exclude that there can exist an interesting class of
(non)--static (anti)--brane solutions for which $\theta=0$ is an admissible 
gauge. In general such solutions will break worldvolume superconformal symmetry.

\bigskip
{\bf Acknowledgements}. 
We are grateful to I. Bandos, P. Claus, R. Kallosh, C. Preitschopf, 
T. Van Proeyen, D. Smith and B. de Wit
 for very stimulating and valuable discussions.
This work was partially supported by the 
European Commission TMR Programme ERBFMPX-CT96-0045 to which the authors 
are associated, and by the INTAS Grants 93-493-EXT and 96-308. D.S. Is grateful
to the Alexander von Humboldt Foundation for providing him with 
a European Research Grant.


\begin{thebibliography}{99}
\bibitem{Maldacena} J. Maldacena, {\it Adv. Theor. Math. Phys.} {\bf 2} 
(1998) 231. 
\bibitem{GKP} S.S~Gubser, I.R.~Klebanov and A.M.~Polyakov, {\it Phys. 
Lett.} {\bf B428} (1998) 105. 
\bibitem{Witten} E.~Witten, {\it Adv. Theor. Math. Phys.} {\bf 2} (1998) 253. 
\bibitem{CKVP} P.~Claus, R.~Kallosh and A.~Van~Proeyen, {\it Nucl. Phys} 
{\bf B518} (1998) 117. 
\bibitem{CKKTVP} P.~Claus, R.~Kallosh, J.~Kumar, P.K.~Townsend and 
A.~Van~Proeyen, {\it JHEP} {\bf 06} (1998) 004. 
\bibitem{MT} R.R.~Metsaev and A.A.~Tseytlin, {\it Type IIB 
Superstring Action in $AdS_5 \times S^5$ Background}, {\tt hep-th/9805028}. 
\bibitem{KRR} R.~Kallosh, J.~Rahmfeld and A.~Rajaraman, {\it Near Horizon 
Superspace}, {\tt hep-th/9805217}.
\bibitem{MT2} R.R.~Metsaev and A.A.~Tseytlin, {\it Type IIB Supersymmetric 
D3 brane action in $adS_5\times S^5$}, {\tt hep-th/9806095}. 
\bibitem{DFFFTT} G.~Dall'Agata, D.~Fabbri, C.~Fraser, P.~Fr\'e, P.~Termonia 
and M.~Trigiante, {\it The $OSp(8|4)$ singleton action from the 
supermembrane}, {\tt hep-th/9807115}. 
\bibitem{RKILL} R.~Kallosh, {\it Superconformal Actions in Killing Gauge}, 
{\tt hep-th/9807206}. 
\bibitem{P} I. Pesando, {\it A k Gauge Fixed Type IIB Superstring Action on 
$AdS_5\times S^5$}, {\tt hep-th/9808020}. 
{\it All roads lead to Rome: Supersolvables and Supercosets,} {\tt 
hep-th/9808146}.
\bibitem{KR} R.~Kallosh and J.~Rahmfeld, {\it The GS String Action on 
$AdS_5\times S^5$}, {\tt hep-th/9808038}. 
\bibitem{dWPPS} B.~de~Wit, K.~Peeters, J.~Plefka and A.~Sevrin, {\it The 
M-theory Two-Brane in $AdS_4\times S^7$ and $AdS_7\times S^4$}, {\tt 
hep-th/9808052}. 
\bibitem{KT} R.~Kallosh and A.A.~Tseytlin, {\it Simplifying superstring 
action on $AdS_5 \times S^5$}, {\tt hep-th/9808088}. 
\bibitem{Claus} P. Claus,{\it Super M-brane actions in $adS_4\times S^7$ and 
$adS_7\times S^4$},
{\tt hep-th/9809045}.
\bibitem{Oda} I. Oda, {\it Super D-string Action on $AdS_5 \times S^5$}, 
{\tt hep-th/9809076}.
\bibitem{RR} J. Rahmfeld and A. Rajaraman, The GS String Action on AdS(3)xS(3)
with Ramond--Ramond Charge, {\tt hep-th/9809164}.
\bibitem{P1} I. Pesando, 
{\it The GS Type IIB Superstring Action on AdS(3) X S(3) X T(4)}, 
{\tt hep-th/9809145}.
\bibitem{BST} E.~Bergshoeff, E.~Sezgin and P.K.~Townsend, {\it Phys. 
Lett.}, {\bf 189B} (1987) 75, {\it Ann. Phys.} {\bf 185} (1988) 330. 
\bibitem{dbrane}
M.Cederwall, A. von Gussich, B.E.W. Nilsson, A. Westerberg,
{\sl Nucl.Phys.} {\bf B490}
(1997) 179;\\
M. Aganagic, C. Popescu, J.H. Schwarz, {\it Phys. Lett.} {\bf B393} (1997) 
311;\\
E. Bergshoeff and P. K. Townsend,
{\it Nucl.Phys.} {\bf B490}
(1997) 145.
\bibitem{5b} I. Bandos, K. Lechner, A. Nurmagambetov, P. Pasti, D. 
Sorokin and M. Tonin, Phys. Rev. Lett. {\bf 78} 
(1997) 4332. \\
M.~Aganagic, J.~Park, C.~Popescu and J.H.~Schwarz, {\it 
Nucl. Phys.} {\bf B496} (1997) 191. 
\bibitem{end} E.~Bergshoeff, M. J. Duff, C. N. Pope and E. Sezgin,
{\it Phys. Lett.} {\bf B199} (1987) 69. 
\bibitem{LPR} H.~L\"u, C.N.~Pope and J.~Rahmfeld, {\it A Construction of Killing 
Spinors on $S^n$}, {\tt hep-th/9805151}. 
\bibitem{abz} V. P. Akulov, I. A. Bandos, V. G. Zima, 
{\it Nonlinear realization of
Extended Superconformal Symmetry}, {\it Theor. Math. Phys.} {\bf 56} (1983)
635.
\bibitem{stv} D. Sorokin, V. Tkach and D. Volkov, {\it Mod. Phys. Lett.}
{\bf A4} (1989) 901.
\bibitem{bpstv} I. Bandos, P. Pasti, D. Sorokin, M.  Tonin and D. Volkov,
{\it Nucl. Phys.} {\bf B446} (1995) 79.
\bibitem{hsw}
P.S. Howe and E. Sezgin, {\sl Phys.  Lett.} 
{\bf B394} (1997) 62.\\
P.S.  Howe, E.  Sezgin and P. C.  West, {\sl Phys.  Lett.}
{\bf B399} (1997) 49.
\bibitem{rev} P. S. Howe, E. Sezgin and P. C. West, 
{\it Aspects of Superembeddings},  Proceedings of the Volkov Memorial Seminar
``Supersymmetry and Quantum Field Theory" (Kharkov, January 5-7, 1997).
Lecture notes in physics, Vol. 509, p. 64. Springer--Verlag, Berlin, Heidelberg,
1998.\\
I. Bandos, P. Pasti, D. Sorokin and M. Tonin, {\it Superbrane Actions and
Geometrical Approach}, {\it Ibid.} p. 79.
\end{thebibliography}
 \end{document}